\documentclass[graybox]{svmult}
\usepackage{etex}
\usepackage{geometry}
\usepackage[document]{ragged2e}
\usepackage{multirow}
\usepackage{changepage}
\usepackage[pdftex]{graphicx}
  \DeclareGraphicsExtensions{.png,.pdf}
\usepackage[utf8]{inputenc}
\usepackage[english]{babel}
\usepackage{lmodern}
\usepackage{makeidx}
\usepackage{textcomp,latexsym}
\usepackage{scrtime}
\usepackage{algorithm}
\usepackage{algpseudocode}
\usepackage{pgfplots}
\usepackage{tikz}
\pgfplotsset{
	scaled ticks = false,
	tick label style={/pgf/number format/fixed, /pgf/number format/precision=4},
}
\pgfplotscreateplotcyclelist{color}{%
	{red,mark=o,mark size=2.5},
	{blue,mark=square,mark size=2.5},
	{black,mark=triangle,mark size=2.5},
	{yellow!80!black,mark=diamond,mark size=2.5},
	{brown,mark=10-pointed star,mark size=2.5},
	{teal,mark=x,mark size=2.5},
	{orange,mark=*,mark size=2},
	{violet,mark=square*,mark size=2},
	{cyan,mark=triangle*,mark size=2},
	{green!70!black,mark=diamond*,mark size=2},
	{magenta,mark=asterisk,mark size=2.5},
	{olive,mark=+,mark size=1.5},
	{lime!90!black,mark=oplus,mark size=2.5},
	{pink!90!black,mark=otimes,mark size=1.5},
	{purple,mark=halfcircle,mark size=1.5},
	{gray,mark=x,mark size=1.5}%
}
\newcommand{\pgfbars}{width=350pt,
	height=100pt,
	legend pos=outer north east,
	xtick=data,
	try min ticks=5,
	y label style={at={(axis description cs:-0.02,0.5)}},
	ybar,
	enlargelimits=0.2}
\usepackage{comment}
\usepackage{textcomp,marvosym}

\usepackage{fixltx2e}
\usepackage{subfigure}
\usepackage{adjustbox}
\usepackage{amsmath,amssymb}

\usepackage{cite}

\usepackage{nameref,hyperref}

\usepackage[right]{lineno}

\usepackage{microtype}
\DisableLigatures[f]{encoding = *, family = * }

\usepackage{rotating}

\raggedright
\usepackage[aboveskip=1pt,labelfont=bf,labelsep=period,justification=raggedright,singlelinecheck=off]{caption}
\usepackage[square,sort,comma,numbers]{natbib}
\makeatletter
\renewcommand{\@biblabel}[1]{\quad#1.}
\makeatother

\date{}

\usepackage{lastpage,fancyhdr,graphicx}
\usepackage{epstopdf}
\makeindex             

\begin{document}

\title*{Investigating Cooperativity of Overlapping Community Structures in Social Networks}
\author{Mohsen Shahriari, Ralf Klamma and Matthias Jarke}
\institute{Mohsen Shahriari, Ralf Klamma and Matthias Jarke \at Chair of Computer Science 5 (Databases \& Information Systems), RWTH Aachen University, Ahornstr. 55, 52056 Aachen, Germany,  \email{{shahriari,klamma,jarke}@dbis.rwth-aachen.de}
}
%
%
\maketitle

\abstract*{ \justify Many real-world networks can be modeled by networks of interacting agents. Analysis of these interactions can reveal fundamental properties from these networks. Estimating the amount of collaboration in a network corresponding to connections in a learning environment can reveal to what extent learners share their experience and knowledge with other learners. Alternatively, analyzing the network of interactions in an open source software project can manifest indicators showing the efficiency of collaborations. One central problem in such domains is the low cooperativity values of networks due to the low cooperativity values of their respective communities. So administrators should not only understand and predict the cooperativity of networks but also they need to evaluate their respective community structures. To approach this issue, in this paper, we address two domains of open source software projects and learning forums. As such, we calculate the amount of cooperativity in the corresponding networks and communities of these domains by applying several community detection algorithms.
Moreover, we investigated the community properties and identified the significant properties for estimating the network and community cooperativity. Correspondingly, we identified to what extent various community detection algorithms affect the identification of significant properties and prediction of cooperativity. We also fabricated binary and regression prediction models using the community properties. Our results and constructed models can be used to infer cooperativity of community structures from their respective properties. When predicting high defective structures in networks, administrators can look for useful drives to increase the collaborations.}

\abstract{Many real-world networks can be modeled by networks of interacting agents. Analysis of these interactions can reveal fundamental properties from these networks. Estimating the amount of collaboration in a network corresponding to connections in a learning environment can reveal to what extent learners share their experience and knowledge with other learners. Alternatively, analyzing the network of interactions in an open source software project can manifest indicators showing the efficiency of collaborations. One central problem in such domains is the low cooperativity values of networks due to the low cooperativity values of their respective communities. So administrators should not only understand and predict the cooperativity of networks but also they need to evaluate their respective community structures. To approach this issue, in this paper, we address two domains of open source software projects and learning forums. As such, we calculate the amount of cooperativity in the corresponding networks and communities of these domains by applying several community detection algorithms.
Moreover, we investigated the community properties and identified the significant properties for estimating the network and community cooperativity. Correspondingly, we identified to what extent various community detection algorithms affect the identification of significant properties and prediction of cooperativity. We also fabricated binary and regression prediction models using the community properties. Our results and constructed models can be used to infer cooperativity of community structures from their respective properties. When predicting high defective structures in networks, administrators can look for useful drives to increase the collaborations.}


\section{Introduction}
\label{Introduction}
Complex networks manifest an abstract representation of complex systems, in which one can model elements of such a system by using nodes as entities and edges as the connections among these nodes. Such a network-based representation is suitable for the modeling and description of systems of different disciplines including biology, sociology, physics and computer science. Researchers have revealed different properties from complex networks including power-law degree distribution, motifs, community structures, shrinking diameter, and so on \cite{LKFa07, MSI*02, PDFV05}. In this regard, researchers can describe online social networks by the analytical and models from complex networks as well as data mining.

Online social networks consist of densely connected components named (overlapping) communities, which the nodes can be the member of more than one community. These (overlapping) community structures can be explicit like a student who is studying at the university and also simultaneously attend a sports club. In such a case, we already consider borders knowing the membership of a person to its corresponding community \cite{JaLe12, PDFV05}. However, we require algorithms for the detection of implicit community structures, which we simultaneously need to evaluate the suitability of these algorithms. Detection and analysis of community structures bring multifaceted benefits to the research and development of network science and social network analysis. In this regard, community structures reveal extra information of node memberships and collective behavior of groups. For instance, network science researchers have developed algorithms for the recommendation of items to users, which use temporal dynamics of community structures \cite{DoJe15, CNZh15}. Besides, link prediction models and ranking algorithms have been developed which use the information revealed by community structures \cite{SPKl15}. Other applications like misbehavior detection, routing and biology and analyzing criminal networks also benefit from implicit/explicit community information \cite{MOTs14, AHHa11, TSTB16, CBPi17}.

Another aspect of network studies relates to the cooperation between selfish individuals that acts as a central part of human society. Conflict situations occur between individual interests and collective interests, which is called social dilemmas \cite{Nowa06}. Usually, social dilemmas have two main properties. First, an agent can obtain its highest benefit by being selfish and choosing non-cooperative behavior. Secondly, the group benefits from cooperative behaviors of individuals \cite{DaMe00}. We can find social dilemmas in various applications. For instance, in learning environments, the groups of learners benefit by sharing the knowledge and helping each other. In such a situation if all the learners avoid sharing the knowledge, then no one will benefit, and it may not be a right situation. Concerning the OSS environments, this is also true, in which someone does not develop the software, but he still can benefit from using it. If all the members stay selfish, the software will face a halt in its development process.

There are several problems in the intersection of cooperativity and community structures; as an example, let us consider a university which consists of various groups working on different topics. If we compute the amount of cooperativity in the whole network as well as their communities, we can use it as a metric to evaluate the productivity and efficiency of the groups and the whole university. Identifying less productive environments and knowing about their low cooperativity can help us to identify these communities and use drives to improve such structures and their internal collaboration and cooperation. Besides, we observe several problems that we can find solutions by considering cooperativity of network and (overlapping) community structures. For instance, in learning environments, some users only play the role of lurkers, which they only observe and benefit from the shared knowledge by others; however, they never contribute and answer any questions. If the number of non-cooperative members in a community increases it will degrade the overall benefit of a community. To improve the efficiency of such an environment, we require not only approaches to evaluate the amount of cooperativity, but also we need to know which properties are the most significant in predicting cooperativity of community structures. 

Researchers have combined game theory approaches and motif discovery techniques to study correlation among motif frequencies and cooperativity level of motifs \cite{SRJa10}. The similar problem can be to compute the correlation between the rank of nodes and their cooperativity level in complex networks, which connect ranking algorithms to game theoretic methods, i.e., Prisoner's Dilemma. Although investigating community properties and predicting community cooperativity can be useful for various applications; to the best of our knowledge, network science has not investigated community cooperativity prediction. Similarly, this problem holds for OSS environments, as it may even be more critical for the aim and progress of the project to predict the amount of community cooperativity. However, we know very few concerning the cooperation \& defection of implicit community structures including learning and OSS networks.

Stating the above challenges, we mention the research question as follows:
\begin{itemize}
\item To what extent do the amount of cooperation \& defection of OSS and LF networks and their community structures differ?
\item How much network \& community properties correlate with the amount of network \& community cooperativity in these two domains?
\item How precise can we predict the amount of network and community properties using structural properties of networks and communities?
\item How much cooperativity of networks and community structures depend on the applied community detection algorithm, and the detected implicit structures?
\end{itemize}

To answer these research questions, we investigate two main domains including LF and OSS development environments by investigating and predicting network and community cooperativities. First, we investigate these domains by computing the amount of cooperativity in their corresponding networks using the Prisoner's Dilemma. Secondly, we examine how network properties favor the cooperativity; moreover, we use prediction models, i.e., classifiers, to construct models for the prediction of cooperativity in these networks. Besides, we detect community structures in these networks and investigate their cooperativity. In this regard, we used different algorithms to understand how the usage of various algorithms can affect the analysis of community cooperativity.
Additionally, we computed properties of detected (overlapping) communities such as size, density, average and standard deviation of node degrees, by using several community detection algorithms. On the other hand, we calculated the cooperativity of every community with evolutionary games, which we finally constructed classifiers for predicting the cooperativity of groups and community structures. We could figure out that the amount of cooperativity in OSS networks and communities is higher than LF networks and communities. Second, community cooperativity shows negative correlation with their properties and properties having high positive or negative correlation with the amount of cooperativity has a significant effect on the prediction of community cooperativity. Moreover, for community detection algorithms which identify communities of small sizes, density has been the important feature for predicting community cooperativity. Also, for community detection algorithms which detect communities with large sizes, size of the community and average degree of nodes in the community have been essential to predict the amount of community cooperativity.

The rest of this article continues as follows: In section \ref{RelatedWork}, we review the related work and mention the contribution of our work. In section \ref{Method}, we explain how to compute the cooperativity on the networks as well as how we detect community structures. Moreover, we describe the network and community cooperativity prediction models. In section \ref{EvaluationProtocol}, we describe the properties of the datasets and the metrics. In section \ref{ResultSection}, we discuss the results of the simulations and the computed predictions. Finally, in section \ref{Conclusions}, we mention the conclusion as well as directions for future works.

\section{Related Work}
\label{RelatedWork}
\label{RelatedWork}
In this section, we review the related work concerning cooperation and defection as well as community detection algorithms separately. We also study cooperativity of community structures and mention the contribution of our work.
 \subsection{Cooperation and Defection}
   \label{relatedworkcooperationdefection}
   
   Study of cooperation and selfish behavior has received much attention recently. Literature identified several mechanisms that favor cooperation. Hamilton \cite{Hami64} proposed the inclusive fitness theory, which is also known as kin selection. Kin selection \cite{Smit64} is based on two fundamental concepts. First, it states that a higher relatedness favors cooperative interactions. The second mentions that the higher payoff obtained by a player's relatives also increases its fitness. Trivers \cite{Triv71} proposed the concept of reciprocal altruism, which states that cooperation occurs through mutual help. By studying the literature, we get to know about other types of reciprocity, i.e., direct and indirect reciprocity, which we can observe in social systems. Nowak proposed a common framework to investigate reciprocity through evolutionary game theory \cite{Nowa06}.
   
   Neumann and Morgenstern \cite{NeMo07} invented game theory as a robust framework to study economic and strategic human decisions. Similarly, Smith and Price \cite{SmPr73} used game theory as a tool to study biological structures. They offered a new perspective on human behavior, which they studied the cooperation between humans through game theory \cite{Axel84, Nowa06}. Axelrod \cite{Axel84}, for the first time, studied the evolution of cooperation, which he investigated the mechanism of direct reciprocity through game theory approaches. Axelrod considered a sequence of strategies for the iterated prisoner's dilemma. He could show that Tit-for-Tat is the most successful strategy, which favors cooperation. Indirect reciprocity was studied by Alexander \cite{Alex87}, which an individual does not need to get an immediate benefit for interactions. In other words, in indirect reciprocity, a player may cooperate with another player to get a benefit from a later interaction with a third player. For instance, in an open source software development environment, developers participate to gain the reputation as well as skills while cooperating with others. Several variants of games used for direct reciprocity have been applied to study indirect reciprocity \cite{NoSi98, BoRi89}.
   
   Network cooperativity studies, on the other hand, investigated the cooperation and defection on a network of nodes connected to each other. As such, Nowak and May studied evolutionary prisoner's dilemma on regular networks \cite{NoMa92}. They showed that evolution of cooperation depends on the structure of the network and regular networks can promote cooperation. There exists other models and studies on network cooperativity \cite{SzFa07, RCSa09, DoHa05, HCSy11}. Scale-free networks were investigated by Santos et al. \cite{SaPa05}, which they showed that preferential attachments in complex networks favor cooperation.

\subsection{Community Detection}
Studies on detection of community structures started by algorithms which detect disjoint communities \cite{ClNM04, LJZQ11}; however, they are not proper to identify communities in real-world networks. The literature on community detection algorithms contains several different categorizations. In a global perspective, we designate the approaches to three classes of static or dynamic, local or global and structural or content-based algorithms. In addition to the above categories, we can nominate algorithms in this area as categories such as clique percolation, line graphs or link communities, local optimization or leader-based methods, random walks or agent-based algorithms. Global methods apply universal metrics to identify communities \cite{ClNM04}. These metrics can be, for example, modularity, and we can optimize them through a global optimization approach \cite{JiMc12, GhLe10}. Local methods of community detection consider local information of the network and thus approaches using random walk processes, cliques, influential nodes or leaders work as local methods \cite{SCCH09, LSHa08, KKKS08, ToSa15, SSKo11, WLLN17, LNDe16}. Furthermore, considering other sources of information may provide more realistic data to identify communities. Content-based and attribute-based approaches take attributes of nodes and edges besides to the connection information \cite{YMLe13, RuFP12}. In addition to locality and relationship information, some algorithms are suitable for temporal environments. Complex networks are dynamic; however, static community detection algorithms can also be applied on each snapshot of the graph separately. For temporal network properties, static solutions may not be suitable and stable, and thus adaptive methods of community detection emerged. Adaptive methods approach the issues encountered in static methods, and they behave more stable \cite{AHSu14, CVRa13}. For a comparison of community detection methods, we refer to \cite{YATe16,FoHr16}.

\subsection{Cooperativity of Community Structures}
 Because studying of community structures is the main focus in this article, we also review cooperation and defection in networks with communities. As such, Luthi et al. \cite{LPTo08} detected community structures on several networks including scale-free, synthetic and real-world social networks. They showed that communities tend to adopt the same structure and thus cooperators get separated from defectors. Another study by Chen et al. \cite{CFWa07} on community structures showed that reducing of inner-community and inter-community links can favor cooperation. They used a model based on Santos et al. \cite{SaPa05} on subgraphs of community structures with different average degrees. Yong-Kui et al. \cite{YZXL09} used evolutionary prisoner's dilemma to show that the higher community sizes results in smaller cooperativity. Salehi et al. \cite{SRJa10} studied cooperation in statistical repeated structures in complex networks, i.e., motifs \cite{SMMA02}. They used a variant of prisoner's dilemma based on replicator dynamic to show that cooperativity correlates with the lower significance of motif structures. We see, therefore, that we still know very few regarding cooperativity of implicit community structures detected by (overlapping) community detection algorithms. Moreover, there is no study to construct classification models using (overlapping) community properties to predict community cooperativity. 
 
\subsection{Contributions of this Work} 

After having identified the research niche in this area, we mention our contributions as follows:
\begin{itemize}
\item We computed cooperativity of LF and OSS networks, and we figured out that OSS networks have a higher level of cooperativity compared to LF networks.
\item We detected community structures in networks of LF and OSS, and we computed the amount of cooperativity of implicit community structures. For the first time, we fabricated binary and regression classifiers to predict the amount of community cooperativity.
\item Both network properties and community properties in LF and OSS networks show negative correlations with the amount of network and community cooperativity, respectively. We later observed that features with the high positive or negative correlation with the amount of cooperativity showed a high impact for the cooperativity prediction.
\item In the network level, clustering coefficient and average degree have the highest importance for predicting cooperativity in both the LF and OSS networks.
\item For predicting community cooperativity, different algorithms can lead to different important properties. For example, for algorithms like SSK and CliZZ which detect smaller communities, size and density are essential for predicting community cooperativity. For algorithms like SLPA, Walktrap, and InfoMap, which detect bigger communities, degree deviation and density are essential properties.
\end{itemize}

\section{Method}
\label{Method}

In this section, we describe our methodology to compute the amount of cooperativity on networks and community structures as well as the applied community detection algorithms. We explain how we analyze community cooperativity, and how we use community properties to construct community cooperativity prediction models.

\subsection{Cooperation and Defection}
\label{CooperationDefection}
Network researchers have been interested in understanding the dynamics of cooperation and selfish behaviors of individuals and groups. We may not separate interests and benefits of groups from individuals and vice verse. In other words, the cooperation of one person with others has reciprocal effects that need to be studied. In fact, the pursuit of selfish agents' behavior might be detrimental for the whole community. These issues have been modeled with game theory approaches such as Prisoner's Dilemma (PD) \cite{RAB*14, RBH*14}. Merrill M. Flood and Melvin Dresher developed this game in around 1950's that has been under much investigation in computational social and computer sciences. There are some variations to this game; however, the original game which is common is a two player game that each agent can take two strategies; either cooperation or non-cooperation. Non-cooperation is also known as defection. The underlying mindset behind this problem is that cooperation of both agents leads to higher lucrative outcomes while defection increases individual benefits \cite{SaPa05}.

The original PD game is the story of two persons imprisoned which the prosecutor cannot convict them based on the existing evidence. Here the court specifies a bonus for the pairs, and they can behave as follows: both prisoners would serve one year in jail if both stay silent. If one of them betrays, he will be free, and the other must serve three years in prison and vice verse. Finally, if both of them betray (defect), they require serving two years in jail. Rationally speaking, pairwise cooperation helps more beneficial outcomes to both players and nonreciprocal defection is useful for each of them \cite{EaKl10}. To simulate the problem, we consider the PD game with a more formal definition. We investigate an evolutionary match played on a network of interactions among individuals or players. Each player can take a set of rational behaviors named strategies. Here, we mention Cooperation(C) and Defection (D) as possible ones. Based on the selected strategy, she will obtain a payoff; higher payoff values are preferred. We denote the payoff by a matrix which Table \ref{PayoffMatrix} shows it. Elements of the payoff matrix are defined as follows:
\begin{itemize}
\item If both players cooperate then they both receive a payoff $R$.
\item If one cooperates and the other defects then the later receives a $T$ and the former obtains $S$.
\item If both players defect then $P$ is granted to both.
\end{itemize}

Each agent plays with its neighbors in the network and receives the corresponding payoff (fitness). With $T > R >P >S$, the non-cooperation dynamic prevails and reaches a Nash Equilibrium. In contrast, the reciprocal cooperation of individuals prevails unilateral cooperation, and thus mutual cooperation generates higher payoff \cite{Gint09}.
\begin{table}[htb!]
\centering
\caption{Payoff matrix for the basic PD game.}
\begin{tabular}{l l l}
\hline
\hline
 & \textbf{C} &  \textbf{D}\\
 \textbf{C} & (R;R) & (S;T) \\
 \textbf{D}  & (T;S) & (P;P) \\

\hline
\end{tabular}

\label{PayoffMatrix}
\end{table}

In general, cooperative $(\#C)$ level of an evolutionary game on a network of agents interacting through PD game is the final number of cooperative agents. As we want to calculate the correlation between node cooperativity and its rank, we require knowing cooperation status of each agent. In this regard, we assign a binary variable as the cooperation status of an agent; 1 denotes cooperation, and 0 shows defection. We start the game with an equal number of cooperators and defectors. Afterwards, in each evolutionary step of the PD game, each agent plays with its neighbors and its payoff value is updated based on the payoff matrix introduced in Table \ref{PayoffMatrix}. We adopt the parameters for the payoff matrix proposed by Nowak and May, and thus we consider $T=b>1, P=S=0, R=1 $ which $T$ is the propensity to defect \cite{NoMa92}. Synchronously, players update their payoffs; however, through the evolutionary steps of the game, players change their strategies. In other words, each player randomly looks at one of its neighbors and change its strategy with a probability given by the equation \ref{ProbFormula} only if its payoff is less than of its neighbor. In fact, regarding node $i$, it compares its payoff with one random neighbor $j$ and changes its strategy with the following probability:
\begin{equation}
\label{ProbFormula}
		P_{i\Rightarrow j} = { PO_j - PO_i \over {b * maximum(\#d_i,\#d_j)} }.
\end{equation} \\
The pseudo code better shows the process. 
\begin{algorithm}[htb!]
\caption{Evolutionary Prisoner's Dilemma Game on Complex Networks}\label{alg:PD-Game}
\begin{algorithmic}[1]
\State $\textit{T} \gets \textit{b}$
\State $\textit{P, S} \gets \textit{0}$
\State $\textit{R} \gets \textit{1}$
\State $\textit{\#Cooperators} \gets N/2$
\State $\textit{\#Defectors} \gets N/2$
\While{\textit{Repetition of the Evolutionary PD Game} $\leqslant$ \textit{Threshold}}
\While{\textit{Iterations} $\leqslant$ \textit{Time Window}}
\For{each node i in the set of all nodes} 
\For{ each node j as neighbor of node i} 
\If{$\textit{strategy(i)} ==C, \textit{strategy(j)==C}$}
\State $\textit{payoffNew(i)} \gets \textit{payoffOld(i)+R}$
\EndIf
\If {$\textit{ strategy(i)==D}, \textit{strategy(j)==C}$}
\State $\textit{payoffNew(i)} \gets \textit{payoffOld(i)+T}$
\EndIf
\EndFor 
\State $\textit{j} \gets \textit{random neighbor of node i}$
\If {$\textit{ payoff(i)}\leqslant \textit{payoff(j)}$}
\State $\textit{strategy(i)} \gets \textit{strategy(j)}$  with $ P_{i\Rightarrow j} = { PO_j - PO_i \over {b * maximum(\#d_i,\#d_j)} }.$
\EndIf
\EndFor

\EndWhile
\State Average over the game realizations
\EndWhile
\Return $\textit{Cooperativitiy Level}$
\end{algorithmic}
\end{algorithm}
\subsection{Community Detection and Evaluation Protocol}
\label{OCD}
   
 To find overlapping communities in several forums, we use community detection algorithms including InfoMap, SLPA, SSK, and CLiZZ. We adopt the commonly-used approach by Nowak and May \cite{NoMa92}. The parameters in this method are $R = 1$, $S = P = 0$ and $T = b > 1$, which this approach can be considered as a simplified version of the Prisoner's Dilemma (PD) with only one parameter. As for update dynamic, we use the replicator dynamic (RP). In RP, updating rule depends on the tendency to imitate others ($b$ parameter) that we set it to 1.5. We consider a population structure, which we consider each node as a player. We randomly initialize the players with strategies, afterward, they play the game against all their neighbors. The payoff a player receives is the aggregation of each payoff through each of the games. Finally, the strategy of all agents is updated based on a global update rule. Here, we run our simulation for at least 1000 iterations. Afterwards, we check whether players have reached a stationary state. Therefore, we compute the standard deviation of the network cooperativity values of the last 200 iterations. We consider the conditions of a stable state as fulfilled whenever the standard deviation of the last 200 iterations is less or equal to $ 1/\sqrt{N} $. We let the system run for another 200 iterations if this is not the case. The simulation run until the system reaches the maximum of 9000 iterations.

 As the system reaches a steady state, we compute the network cooperativity and the agents' cooperativities for the simulation. We calculated the network cooperativity as the average cooperativity value of the network over the last 200 iterations. Also, we considered the cooperativity value for an agent, which it can be computed based on the number of time the agent played cooperation within the last 200 iterations divided by 200. To prevent biases in our simulations, we run every simulation 200 times and compute the average cooperativity values. The cooperativity $\Delta(N)$ of Network N is the average network cooperativity of the 200 simulations. The same condition is met for agents' cooperativity. Similarly, we calculate cooperativity value of a community. The cooperativity value of a community $C$ is the average cooperativity of its members, in other words, we can compute it as:

   $$\Delta(C) =  \sum_{n \in C}  \frac{ \Delta{n} }{|C|}. $$
   
   We classify a community as cooperative if $\Delta(C) > 0.5$, otherwise as defective. To detect communities, we use one of the (overlapping) community detection algorithms, i.e., DMID, SLPA, etc.   
\subsection{Network and Community Cooperativity Classification Models}
To predict the cooperativity of networks and communities, we used several classifiers. The target variable can be either binary or continuous. Regaring the regression case, we use the computed cooperativity of networks or communities; however, concerning the binary case, we map them to zero and one. For binary prediction, we use a tree and Logistic Regression classifer, and for the regression case we use the Ridge classifier, which we describe them in the following.
\subsubsection{Logistic Regression}
Logistic regression uses a set of features for training and makes a probabilistic model out of the training data. The logistic regression utilizes a cost function to find the parameters of the model. The cost function is as follows:

\begin{equation}
		J(h_\theta(x),y)={{{1 \over 2}\times({{1\over {{{1+\exp(-\theta^{T}x)}}}}-y)}}^{2}}.
		\label{LogisticRegression}
\end{equation}
Where $J$ is the error in the model, $x$ is the vector of features, $\theta$ is the vector of parameters which should be computed by an optimization approach. Finally, $y$ contains the real class values. In our case, it is a vector of +1 and -1 values  \cite{Bish06}. 

\subsubsection{Extremely Randomized Trees (ExtraTreesClassifier)}
Another classifier which we used to predict the binary cooperativity values is ExtraTreesClassifier. Geurts et al. \cite{GEDW06} proposed ExtraTreesClassifier to construct an ensemble of the unpruned decision or regression trees. It splits the vertices based on selecting cut-points randomly, and it applies the whole training samples to fabricate the trees. ExtraTreesClassifier aggregates the predictions resulted from the trees to compute the final prediction by majority vote in case of a classification problem. The benefits of using ExtraTreesClassifier reduces the burden of creating multiple copies of the learning sample by selecting a cut-point at random.
Moreover, choosing the explicit randomization of the cut-point helps to reduce variance more than other weak randomization strategies. Also, using the full training sample to grow the tree helps in minimizing the bias. The authors showed that this classifier leads to higher accuracies, and it is implemented in Scikitlearn library \footnote{\url{http://scikit-learn.org/stable/modules/generated/sklearn.ensemble.ExtraTreesClassifier.html}}.
\subsubsection{Ridge Regression} 
Tikhonov Regularization is also known as Ridge regression, which is primarily a regression algorithm and is applied on problems which do not have a unique solution, which the number of training examples might as well be limited. We have applied the Ridge classifier from the Scikitlearn library \footnote{\url{http://scikit-learn.org/stable/modules/generated/sklearn.linear_model.RidgeClassifier.html}} to construct a regression model for community cooperativity prediction based on community properties. The Ridge classifier consists of a cost function as follows:
\begin{equation}
min(||Y-X(\theta)||^2_2 + \lambda||\theta||^2_2).
\end{equation}

Which $\lambda$ is the term controlling the penalty, $\theta$ is the parameters of the Ridge classifier which is computed by L2 regularization. To identify $\lambda$ value, it uses cross-validation on the training examples. Ridge classifier is supposed to lead into less error as well as lower bias-variance.

\subsection{Overview of the Model}
\begin{figure}[htb!]
\centering

\includegraphics[scale=0.4]{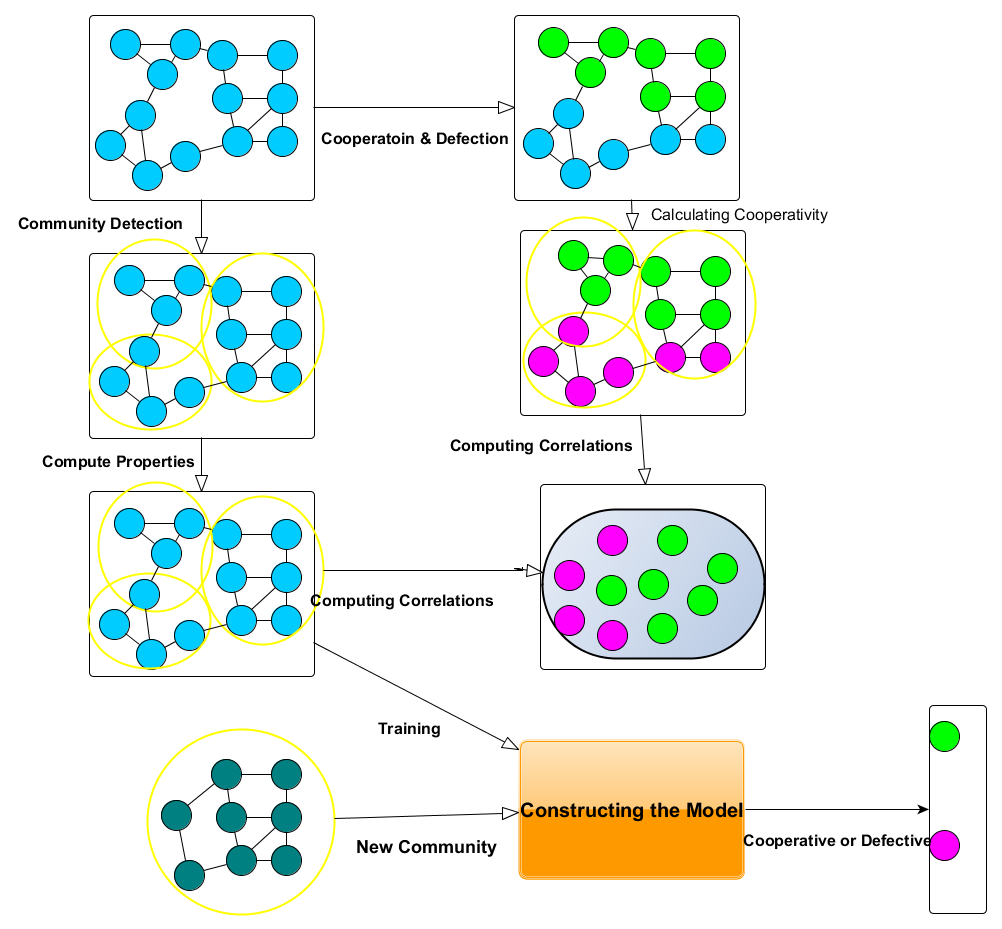}

\caption{This figure shows the applied steps of our approach to analyze and predict the community cooperativities. The green and magenta colors show the cooperative and defective status of a node.}
\label{Methodology}
\end{figure}

To better clarify the applied method for the analysis and prediction of community cooperativities, we refer to Figure \ref{Methodology}. As we can see, on the one side, we detect implicit community structures by applying different community detection algorithms. Simultaneously, we calculate the cooperativity in the respective communities and compute the correlations. Finally, we construct prediction models to predict the cooperativity of a new coming community.  

\section{Datasets and Metrics}
In this section, we describe the applied datasets as well as the used evaluation metrics.
\label{EvaluationProtocol}
\subsection{Datasets}
We performed our simulations on several real-world networks including open source software development (OSS) and learning forum (LF) networks. We interpret each dataset as a simple graph, which we removed edges as well as loops. The updating dynamic of the cooperativity simulations only considered nodes that share one neighbor; therefore we removed every node without an incident edge.

The first data resource comes from open source software projects of the open Bioinformatics foundation. These datasets include JMOL, BioJava, BioPerl and BioPython that were gathered during the development process of their corresponding software. In this regard, the JMOL dataset corresponds to a project that developed a Java-based molecular viewer for chemical structures. The other domains relate to processing of biological data in programming languages including Python, Java, and Perl. We constructed the networks based on the communication threads in these mailing lists, in which we created a node for every participant who sent or received an email. An edge is created when two participants shared an email contact for at least one time. For the creation of the networks over time, we considered release periods from these domains and considered all the emails which correspond to a specific time step.

Concerning the LF networks, we used URCH and STDOCTOR datasets. The URCH dataset is related to the English language learners that discuss issues related to exams such as e.g. GRE; however, STDOCTOR is a platform to discuss medical topics. The original URCH dataset was crawled from the year 2001 to 2010; however, we constructed the monthly networks of the year 2003 to 2005. In this regard, we fabricated an edge between two users that participated in the same thread once during each time step. We as well constructed the network of STDOCTOR corresponding to the year 2009. 

Table \ref{TableNetworks} shows necessary information about the networks of these two domains. In total, we obtain 70 OSS and 48 LF networks, which we showed their respective network properties in this table. This table also shows necessary information concerning, average number of nodes and edges, average degrees, average standard deviation of degrees as well as average clustering coefficient. As we can see in this table, these properties in LF networks are bigger than OSS networks. For instance, concerning average clustering coefficient, we see larger values for LF networks. This observation is as well correct concerning the average degree of these networks.

     \begin{table}[h]
   	\caption{This table shows the basic statistics concerning the OSS and LF networks. Lf networks contain 48 separate networks over the monthly analysis of STDOCTOR and URCH datasets, and OSS networks comprise 70 networks corresponding to the releases of the four different OSS projects.}
   	\label{TableNetworks}
   	\centering
   	\begin{adjustbox}{max width=\textwidth}
   		\begin{tabular}{lcrrrrrr}
   			Name&networks&avg n&avg m   &avg D   &avg deg&avg std deg&avg CC\\ \hline
   			OSS & 70\\  \hline
   			JMOL Releases     &13&123,0   &328,7&0,0511 &4,5801&11,3185&0,4592\\ 
   			BioJava Releases  &10&156,6&250,2&0,0242 &3,0780&5,4755 &0,0710\\ 
   			BioPerl Releases  &18&311,0   &609,4&0,0191 &3,6599&7,4670 &0,1029\\
   			BioPython Releases&29&91,0&133,4&0,0439 &2,8773&4,0441 &0,0280\\ \hline
   			Learning Forum & 48 \\  \hline
   			URCH 2003 &12&235,3&938,7&0,0348&7,7599&10,5915  &0,5652\\ 
   			URCH 2004 &12&289,7&1670,6&0,0350&10,0580&12,6486&0,6008\\	
   			URCH 2005 &12&634,7&5744,8&0,0279&17,5036&23,4578&0,6537\\ 
   			STDOCTOR 2009 &12&309,0&2088,0&0,0432&13,3446&13,3514&0,8694\\ \hline
   			
   		\end{tabular}
   \end{adjustbox}\end{table}
 
\subsection{Evaluation Metrics} 

\subsubsection{Spearman Correlation}
Spearman's rank correlation does not assume whether the relationship among two variables $x$ and $y$ is linear. Moreover, it does not suppose any predefined frequency distributions for two variables. The Spearman measure can be computed as follows:
\begin{equation}
    \rho= 1 - {{ 6 \Sigma d_i^2} \over {n(n^2-1)}},
\end{equation}
which $d_i$ is the difference between the corresponding value of two elements $x$ and $y$, and $n$ is the number of pairs of values \cite{KuVa10}.

\subsubsection{Root Mean Square Error (RMSE)}

Another metric to evaluate the error, which penalizes large values of errors is RMSE \cite{Bish06, TSKu05}. RMSE averages over the square values of errors and can be computed as follows:
 \begin{equation}
RMSE=\sqrt{\frac{\sum_{i=0}^n {\left| f_i-y_i\right|}^2}{n}}.
\label{RMSE}
\end{equation}
\subsubsection{Prediction Accuracy}
While having a classification problem, known classes are binary or multi-target cases. To evaluate such supervised learning problems, one needs to calculate true positive and false positive rates. Positive or negative are signs of the classifier's expectations and better to be named predictions \cite{Bish06}. In contrast, true and false prove to be correspondent with ground-truth labels. If one denotes the true positive rates with (TP) and false positive rates with (FP), then precision can be calculated as follows:
\begin{equation}
Precision= {TP \over TP + FP}.
\label{Precision}
\end{equation}

While having two distinct labels or classes in a binary classification problem, the Prediction Accuracy (PA) may be a better measure indicating the goodness of the prediction as follows:
\begin{equation}
PA= {TP + TN \over TP + TN + FP + FN}.
\label{PredictionAccuracy}
\end{equation}

\section{Results}
\label{ResultSection}
In this section, we discuss the results of the experiment concerning investigation and prediction of community cooperativity.
\subsection{Temptation to Defect and Network Cooperativity}
\begin{figure}[htb!]
\centering

\includegraphics[scale=0.45]{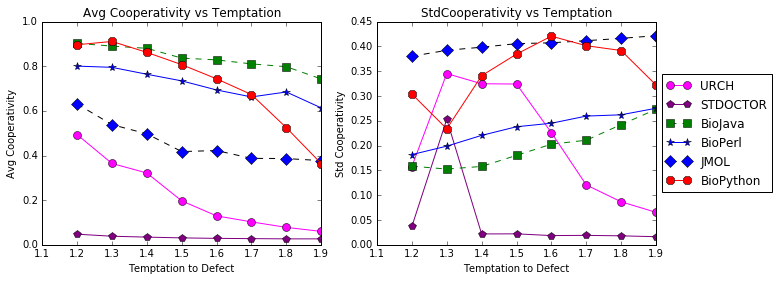}

\caption{This diagram shows the average and standard deviation of cooperativity in OSS and LF networks versus the temptation to defect parameter.}
\label{TemptationToDefect}
\end{figure}

As for prisoner's dilemma with the replicator dynamic, we changed the game Parameter $T$, which is the temptation to defect, between 1.2 and 1.9. It means as we increase the temptation for defection, we expect less cooperativity in the test networks. However, we want to figure out to what extent $T$ parameter affects on cooperativity of learning and OSS forums. As such, we can see in Figure \ref{TemptationToDefect}, information such as average and standard deviation of cooperativity. As we can see average cooperativity decreases for all the networks except STDOCTOR; however, it has a meager amount of cooperativity compared to other networks. When the $T$ parameter approaches 2, then STDOCTOR and URCH have almost equal cooperativity values. Overall, we can observe higher cooperativity values for OSS networks, i.e., BioJava, BioPerl, and BioPython, compared to LF networks, i.e., STDOCTOR and URCH. Concerning OSS networks, BioPython, BioPerl, and BioJava have a higher level of cooperativity compared to JMOL network.

Concerning standard deviation of network cooperativity we can observe that STDOCTOR has almost very low std values near to zero except for $T$=1.3, which shows that the variance of network cooperativity is low. The OSS networks also have the higher variance of network cooperativity, and it increases when $T$ goes up. On the other hand, the variance for URCH first increases and then decreases for higher values of $T$. Overall. OSS networks show a higher variance of cooperativity compared to LF networks except for URCH which has a high variance of cooperativity for smaller $T$ values.

\subsection{Correlations of Network Properties and Network Cooperativity}
   
\begin{figure}[htb!]
\centering

\subfigure{\includegraphics[scale=0.4]{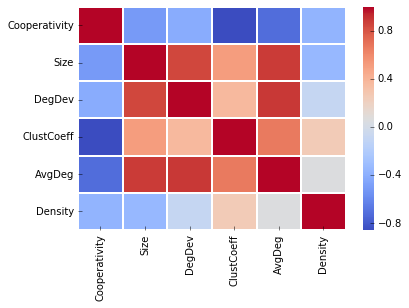}}
\subfigure{\includegraphics[scale=0.4]{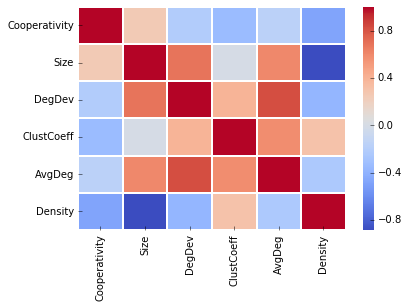} }

\caption{This figure shows the heat map corresponding to the Spearman correlations of the properties. The left heatmap belongs to the learning forums, and the right one belongs to the OSS networks.}
\label{heatmapNetworks}
\end{figure}

We computed the correlations between network properties and their cooperativity values. We considered 70 OSS and 48 LF networks given in Table \ref{TableNetworks}, which each of these networks belong to a specific time span. We used Spearman correlation to compute the correlation values between network properties and their cooperativity values. As for network properties, we used size, density (D), average degrees (AvgDeg), standard deviations (DegDev) of degrees and clustering coefficient (ClustCoeff) values. Figure \ref{heatmapNetworks} shows correlation values for both the Learning forums (left figure) and OSS (right figure). As we can observe, cooperativity has a negative correlation with all the properties in LF networks, in which clustering coefficient followed by average degree have the most negative correlations with the cooperativity values.
Furthermore, in LF networks, we can also see that the properties have some positive correlations together, in which size, degree deviation, and average degree have the highest correlations among each other; however, clustering coefficient and density have lower correlation values with the other properties. Concerning OSS networks, we observe a similar correlation property, in which cooperativity have a negative correlation with the properties of the network including DegDev, ClustCoeff, AvgDeg, and Density; however, size has a small positive correlation with cooperativity values. Besides, size, DegDev, and AvgDeg have the highest correlation values among themselves; however, clustering coefficients and density have fewer correlations compared to them. This observation helps us to get a better perspective and understanding of useful properties for predicting the number of network cooperativities.

\begin{figure}[htb!]
\centering

\subfigure{\includegraphics[scale=0.4]{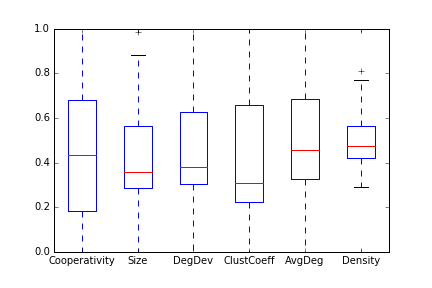}}
\subfigure{\includegraphics[scale=0.4]{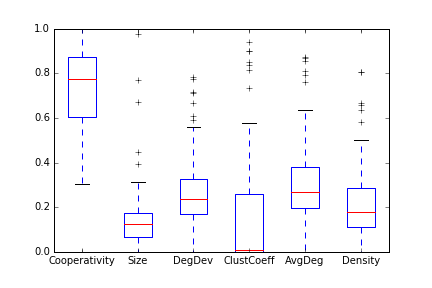} }

\caption{This figure shows the normalized cooperativity values together with normalized network property values. The left plot belongs to Learning Forums (LF), and the right one belongs to the OSS networks.}
\label{OSSLFNetworksBoxPlot}
\end{figure}

To obtain a clear view of the correlations, properties, and the cooperativity values, we plotted the boxplot corresponding to them. We show these visualizations in Figure \ref{OSSLFNetworksBoxPlot}. First, we can observe that OSS networks have a higher cooperativity value compared to the learning forums. Moreover, we can observe the properties in the OSS forums have lower values compared to the learning forums. For instance, clustering coefficient in OSS networks are close to zero; however, clustering coefficient in learning forums are higher. This observation is as well true for other properties like size, density, average and standard deviation of degrees, in which OSS forums have smaller property values compared to the Learning Forums. Altogether, we observe that learning forums are bigger than the OSS networks, and they are more dense with higher values of clustering coefficients. In learning forums, average the degree of nodes is higher than the OSS networks. These observations give us a rather deep understanding of the network properties in these two different domains.

\subsection{Importance of Network Properties in Predicting Network Cooperativity}
\begin{figure}[htb!]
\centering

\subfigure{\includegraphics[scale=0.5]{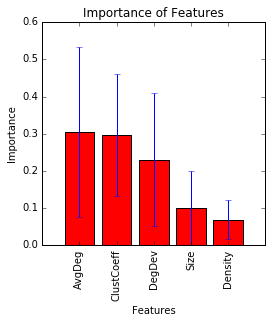}}
\subfigure{\includegraphics[scale=0.5]{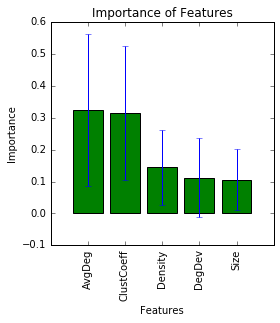} }

\caption{This figure shows the importance of figures in predicting cooperativity values of OSS and Learning forums. The red bars relate to LF, and the green bars relate to OSS networks. Looking at average degree and clustering coefficient can in both domains help us predicting the amount of cooperation in LF and OSS networks.}
\label{FeatureImportanceOSS_LF}
\end{figure}

To compare the importance of features, we use them in prediction models, i.e., predicting cooperativity in networks. We have mapped the cooperativity values to binary values, in which values more prominent than the mean values of cooperativity correspond to one, and less significant ones correspond to zero. To have more stable results, we use the Extra Trees Classifier from the ScikitLearn python library. After fitting the model to the data, we extract the feature importance, in which we show in Figure \ref{FeatureImportanceOSS_LF}. In both datasets, clustering coefficient and average degree had the highest importance for predicting the binary values of cooperativity. Although the density values for OSS networks is lower than LF networks, they are more effective compared to the OSS networks. Based on the box plots from the previous section, we expected clustering coefficient to be useful for prediction of cooperativity; however, average degree as well appeared to be the most successful property among all. Moreover, there are small differences between the importance of features for OSS and LF networks for degree deviation, size, and density, which have the lowest effect in predicting cooperativity in OSS and LF networks. 

\subsection{Importance of Community Properties in Prediction of Community Cooperativity}

\begin{figure}[t!] 
	\centering

		\begin{tikzpicture}[scale=0.8]
	\begin{axis}[\pgfbars,
	x=1.8cm,
	ylabel=CliZZ,
	symbolic x coords={Property}]
	\addplot coordinates {(Property,0.423)};
	\addlegendentry{Density}
	\addplot coordinates {(Property,0.285)};
	\addlegendentry{AvgDev}

	\addplot coordinates {(Property,0.149)};
	\addlegendentry{DegDev}

	\addplot coordinates {(Property,0.142)};
	\addlegendentry{Size}

	\end{axis}
	\end{tikzpicture}
	\begin{tikzpicture}[scale=0.8]
	\begin{axis}[\pgfbars,
	x=1.8cm,
	ylabel=InfoMap,
	symbolic x coords={Property}]
	\addplot coordinates {(Property,0.189)};
	\addlegendentry{Density}
	\addplot coordinates {(Property,0.353)};
	\addlegendentry{AvgDev}

	\addplot coordinates {(Property,0.208)};
	\addlegendentry{DegDev}

	\addplot coordinates {(Property,0.248)};
	\addlegendentry{Size}

	\end{axis}
	\end{tikzpicture}

	\begin{tikzpicture}[scale=0.8]
	\begin{axis}[\pgfbars,
	x=1.8cm,
	ylabel=SLPA,
	symbolic x coords={Property}]
	\addplot coordinates {(Property,0.148)};
	\addlegendentry{Density}
	\addplot coordinates {(Property,0.443)};
	\addlegendentry{AvgDev}

	\addplot coordinates {(Property,0.090)};
	\addlegendentry{DegDev}

	\addplot coordinates {(Property,0.317)};
	\addlegendentry{Size}

	\end{axis}
	\end{tikzpicture}
	\begin{tikzpicture}[scale=0.8]
	\begin{axis}[\pgfbars,
	x=1.8cm,
	ylabel=SSK,
	symbolic x coords={Property}]
	\addplot coordinates {(Property,0.396)};
	\addlegendentry{Density}
	\addplot coordinates {(Property,0.317)};
	\addlegendentry{AvgDev}

	\addplot coordinates {(Property,0.149)};
	\addlegendentry{DegDev}

	\addplot coordinates {(Property,0.137)};
	\addlegendentry{Size}

	\end{axis}
	\end{tikzpicture}
	\begin{tikzpicture}[scale=0.8]
	\begin{axis}[\pgfbars,
	x=1.8cm,
	ylabel=Walktrap,
	symbolic x coords={Property}]
	\addplot coordinates {(Property,0.159)};
	\addlegendentry{Density}
	\addplot coordinates {(Property,0.442)};
	\addlegendentry{AvgDev}

	\addplot coordinates {(Property,0.143)};
	\addlegendentry{DegDev}

	\addplot coordinates {(Property,0.256)};
	\addlegendentry{Size}

	\end{axis}
	\end{tikzpicture}
	
	\caption{This figure shows the importance of community properties for different community detection algorithms on Forum networks. The y axis shows the importance of an corresponding community detection algorithm and the x axis shows the properties.}
	\label{CommunityPropertyImportanceLF}
\end{figure}
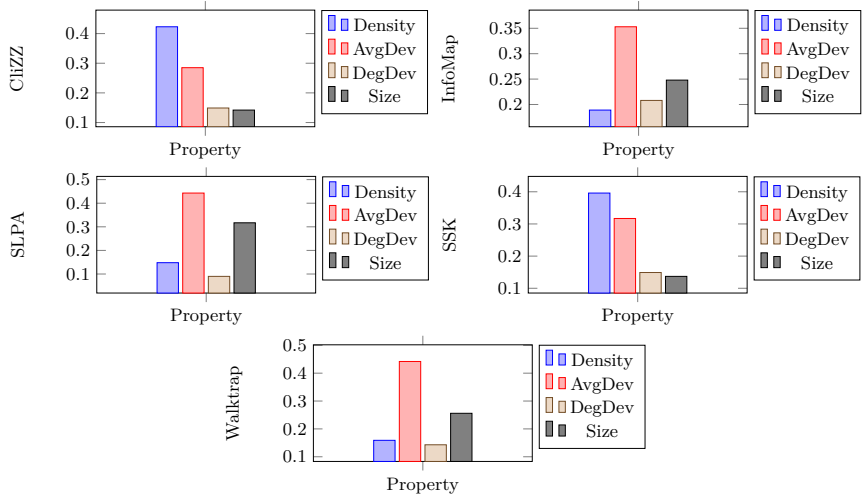

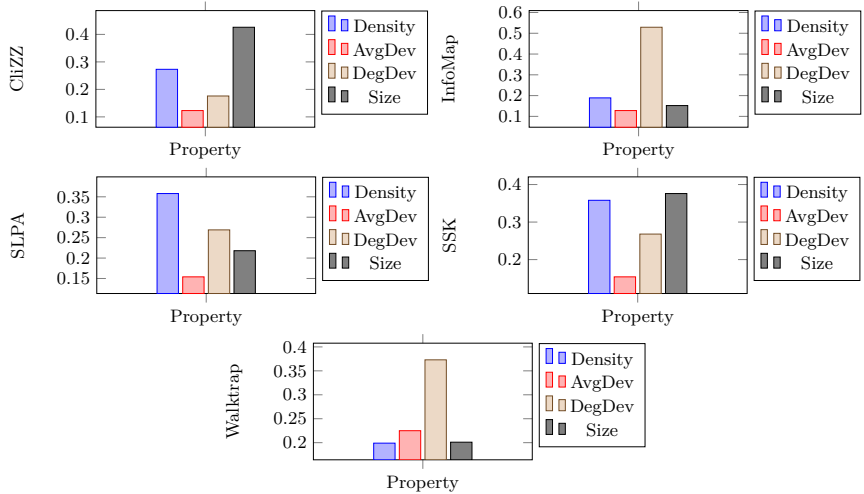
\begin{figure}[t!] 
	\centering

		\begin{tikzpicture}[scale=0.8]
	\begin{axis}[\pgfbars,
	x=1.8cm,
	ylabel=CliZZ,
	symbolic x coords={Property}]
	\addplot coordinates {(Property,0.273)};
	\addlegendentry{Density}
	\addplot coordinates {(Property,0.123)};
	\addlegendentry{AvgDev}

	\addplot coordinates {(Property,0.176)};
	\addlegendentry{DegDev}

	\addplot coordinates {(Property,0.426)};
	\addlegendentry{Size}

	\end{axis}
	\end{tikzpicture}
	\begin{tikzpicture}[scale=0.8]
	\begin{axis}[\pgfbars,
	x=1.8cm,
	ylabel=InfoMap,
	symbolic x coords={Property}]
	\addplot coordinates {(Property,0.189)};
	\addlegendentry{Density}
	\addplot coordinates {(Property,0.128)};
	\addlegendentry{AvgDev}

	\addplot coordinates {(Property,0.529)};
	\addlegendentry{DegDev}

	\addplot coordinates {(Property,0.152)};
	\addlegendentry{Size}

	\end{axis}
	\end{tikzpicture}

	\begin{tikzpicture}[scale=0.8]
	\begin{axis}[\pgfbars,
	x=1.8cm,
	ylabel=SLPA,
	symbolic x coords={Property}]
	\addplot coordinates {(Property,0.358)};
	\addlegendentry{Density}
	\addplot coordinates {(Property,0.154)};
	\addlegendentry{AvgDev}

	\addplot coordinates {(Property,0.269)};
	\addlegendentry{DegDev}

	\addplot coordinates {(Property,0.218)};
	\addlegendentry{Size}

	\end{axis}
	\end{tikzpicture}
	\begin{tikzpicture}[scale=0.8]
	\begin{axis}[\pgfbars,
	x=1.8cm,
	ylabel=SSK,
	symbolic x coords={Property}]
	\addplot coordinates {(Property,0.358)};
	\addlegendentry{Density}
	\addplot coordinates {(Property,0.154)};
	\addlegendentry{AvgDev}

	\addplot coordinates {(Property,0.268)};
	\addlegendentry{DegDev}

	\addplot coordinates {(Property,0.376)};
	\addlegendentry{Size}

	\end{axis}
	\end{tikzpicture}
	\begin{tikzpicture}[scale=0.8]
	\begin{axis}[\pgfbars,
	x=1.8cm,
	ylabel=Walktrap,
	symbolic x coords={Property}]
	\addplot coordinates {(Property,0.199)};
	\addlegendentry{Density}
	\addplot coordinates {(Property,0.225)};
	\addlegendentry{AvgDev}

	\addplot coordinates {(Property,0.373)};
	\addlegendentry{DegDev}

	\addplot coordinates {(Property,0.201)};
	\addlegendentry{Size}

	\end{axis}
	\end{tikzpicture}
	
	\caption{This figure shows the importance of community properties for different community detection algorithms on OSS networks. The y axis shows the importance of an corresponding community detection algorithm and the x axis shows the properties.}
	\label{CommunityPropertiesImportanceOSS}
\end{figure}

Similar to the network cooperativity, we used the Extra Trees Classifier from the ScikitLearn Python library to extract the importance of features in predicting cooperation and the defection of communities. In this regard, we applied different community detection algorithms including CliZZ, SSK, SLPA, InfoMap, and Walktrap. Each of these algorithms detects different communities of various sizes and properties. The bar plot in Figures \ref{CommunityPropertyImportanceLF} and \ref{CommunityPropertiesImportanceOSS} show the importance of properties in the prediction task. In these figures, the y-axis represents the importance of figures identified by using the Extra trees classifier, and the x-axis shows the applied property. The importance of various properties differs when applying different community detection algorithms. In Figure \ref{CommunityPropertyImportanceLF}, we can observe that CliZZ and SSK have similar feature properties, in which density of communities followed by the average degree, degree deviation and size of communities respectively are important in predicting the cooperativity of communities. We can expect this because by looking at Table \ref{CommunityDetectionCorrelations}, we figure out that average degree and density have the highest negative correlation with the amount of cooperativity.

Moreover, SSK and CliZZ instead detect many smaller communities, in which their size does not have a significant effect on the prediction of cooperativity, but rather their density plays an important role. On the other hand, SLPA, InfoMap, and Walktrap have a similar degree of importance for the community properties, which average degree is the most important property. Next, size has the highest effect after average degree. If we look at Table \ref{CommunityDetectionCorrelations}, we figure out that average degree and size have a higher correlation with cooperativity of communities, in which play a more important role compared to SSK and CLiZZ properties. SLPA, InfoMap, and Walktrap tend to detect more similar number and size of communities, which lead to the similar importance of features for the prediction. 

\begin{table}[htb!]
\caption[Correlations of community properties and their cooperativity values using InfoMap, SSK, CliZZ, SLPA and Walktrap algorithms]{This table contains information regarding Spearman correlations of community properties and their cooperativity values using the InfoMap, SSK, CliZZ, SLPA and Walktrap algorithms. The values are computed based on the 70 release-based OSS networks as well as 48 monthly-based learning forums. }
\label{CommunityDetectionCorrelations}
\centering\begin{adjustbox}{max width=\textwidth}\begin{tabular}{*{6}{l}}
\hline
Measure&SSK&CliZZ&SLPA&Walktrap & InfoMap\\ \hline
\multicolumn{2}{l}{Release Networks} &\multicolumn{2}{l}{} \\ \hline
size&-0.373&0.456&0.358&0.056 & 0.185\\
density& 0.396&0.529&-0.373&-0.056 & -0.186\\
avg Deg&-0.165&-0.096&0.182&0.055 & 0.175\\
std Deg&-0.160&-0.249&0.918&0.062 & 0.233\\\hline
\multicolumn{2}{l}{Learning Forums}& \multicolumn{2}{l}{} \\ \hline
size&0.022&0.053&-0.410&-.0357 & -0.225\\
density&-0.544&-0.634&-0.088&-0.048 & -0.135\\
avg Deg&-0.677&-0.649&-0.550&-0.439& -0.392\\
std Deg & -0.335 & -0.301 & -0.025& 0.016 & 0.059\\\hline
\end{tabular}\end{adjustbox}\end{table} 

Figure \ref{CommunityPropertiesImportanceOSS} shows information concerning OSS networks. Similarly, CliZZ and SSK have similar behavior for the importance of features; however, size and then density have the most differentiating effect. By looking at Table \ref{CommunityDetectionCorrelations}, we figure out that these two properties have the highest and lowest Spearman correlation with cooperativity. On the other hand, the results for InfoMap, Walktrap, and SLPA are a bit different than the LF networks, in which degree deviation is the most important property for InfoMap. Concerning SLPA, density and degree deviation seems to have the significant effect in predicting cooperativity, respectively. For Walktrap, degree deviation has the most significant effect. By looking at Table \ref{CommunityDetectionCorrelations}, we can observe that for InfoMap, degree deviation has the highest correlation and density as the second important property has the most negative correlation. Also for SLPA, we can observe the lowest correlation for density and the highest for the deg deviation. This observation affirms that the properties with the highest and lowest correlation with the cooperativity have the highest effect in the prediction of community cooperativity.

\subsection{Predicting Cooperation and Defection of Community Structures}
In this section, we use two approaches for predicting cooperativity of community structures. In the first approach, we map the continuous values of cooperativity, which are between zero and one to either one or zero. We use the mean values of cooperativity as a threshold for the mapping. This way, we obtain binary values for the amount of cooperativity, and we consider communities as either cooperative or defective. In the second approach of the prediction models, we consider the computed continuous cooperative values as the target variable, and we perform a regression analysis using the Ridge regression. In the following, we describe the results conducted by each of these methods.
\subsubsection{Prediction of Community Cooperation and Defection}
For predicting the binary values of cooperativity, i.e., cooperation or defection, we can use different classifiers. Here, we select the logistic regression classifier and the Extra trees classifier, which we as well used for identifying the importance of features. We use the same set of parameters from the previous section. As we can observe in Table \ref{BinaryPredictionForest}, we have applied the Extra Trees classifier to networks of learning forum (LF) as well as the open source software developer networks (OSS). We report the prediction accuracy as well as the mean squared error. Here, SSK and CliZZ achieved the highest prediction accuracy among other algorithms. This is true for both cases including SSK and CliZZ. The prediction accuracies are followed by respectively SLPA, Walktrap and InfoMap.

Similarly, the same pattern holds for the OSS networks, which SSK and CliZZ correctly identify the cooperative as well as the defective classes; moreover, SLPA, Walktrap, and InfoMap obtain lower prediction accuracies. Because we use various estimators in the Extra trees classifier, we also look at the logistic regression classifier. Results of community cooperativity prediction shown in Table \ref{BinaryPredictionLogisticRegression} shows that SSK and CliZZ almost correctly predict cooperation and defection of community structures; however, the prediction accuracy for the learning forums are more stable. Concerning other algorithms, i.e., SLPA, Walktrap, and InfoMap place themselves respectively in other positions. For instance, InfoMap obtains the least prediction accuracy compared to other algorithms.
\begin{table}[t]
	\centering
	\caption{Accuracy and error of community cooperativity prediction using structural community properties and the Extra trees classifier. }
	
		\begin{tabular}{l l l l l l l l }
				\hline \textbf{DS} & \textbf{Metric}  & SSK & CliZZ & SLPA & Walktrap & InfoMap \\
				\hline \parbox[t]{2mm}{\multirow{2}{*}{\rotatebox[origin=c]{90}{LF}}}
				
				 & Prediction Accuracy& 1.0 & 1.0 & 0.913& 0.813& 0.871 \\
				 	
				\cline{2-7}	& MSE    & 0.0   & 0.0 & 0.087 & 0.187 & 0.129 \\ 
			
				\hline
				\parbox[t]{2mm}{\multirow{2}{*}{\rotatebox[origin=c]{90}{OSS}}} 
			        & Prediction Accuracy & 1.0 & 1.0 & 0.731& 0.615 & 0.680\\ 
					\cline{2-7} & MSE & 0.0   & 0.0 & 0.268 & 0.384 & 0.319\\ 
				
				\hline
				
			\end{tabular}
	
	\label{BinaryPredictionForest}
\end{table}

\begin{table}[t]
	\centering
	\caption{Accuracy and error of community cooperativity prediction using structural community properties and the logistic regression classifier. }
	
		\begin{tabular}{l l l l l l l  }
				\hline \textbf{DS} & \textbf{Metric}  & SSK & CliZZ & SLPA & Walktrap & InfoMap \\
				\hline \parbox[t]{2mm}{\multirow{2}{*}{\rotatebox[origin=c]{90}{LF}}}
				
				 & Prediction Accuracy & 1.0 & 0.995 & 0.849& 0.740& 0.736\\

				\cline{2-7}	& MSE     & 0.0  & 0.005 & 0.151 & 0.260 & 0.264 \\ 
			
				\hline
				\parbox[t]{2mm}{\multirow{2}{*}{\rotatebox[origin=c]{90}{OSS}}} 
			        & Prediction Accuracy & 0.657 & 0.828 & 0.619& 0.529& 0.632\\ 
					\cline{2-7} & MSE & 0.342   & 0.172 & 0.381 & 0.471 & 0.368\\ 
				
				\hline
				
			\end{tabular}
	
	\label{BinaryPredictionLogisticRegression}
\end{table}

\subsubsection{Prediction of Continuous Community Cooperativity}
To observe the predictive power of the community detection algorithms, we also predicted the actual continuous values of cooperativity, which we indicate in Table \ref{RegresssionPredictionLinear}. In other words, the average cooperativity obtained from running the prisoner dilemma game is used for prediction of cooperativity using a Ridge regression from the ScikitLearn library. Here, we have reported the MSE and RMSE values. Compliant with the accuracy and error values reported for the binary classification problem, we have a similar pattern concerning the performance of community detection algorithms in the community cooperativity prediction. Similarly, SSK and CliZZ obtain the lowest values of errors in the regression task. Besides, SLPA, Walktrap, and InfoMap obtain other positions.
Therefore, we can conclude that by using the implicit community properties of SSK and CliZZ algorithm, we achieve higher levels of prediction accuracy, i.e., we can more  precisely predict cooperation and defection of communities.

\begin{table}[t]
	\centering
	\caption{Error values using the linear regression classifier for predicting the continuous cooperativity values. }
	
		\begin{tabular}{l l l l l l l  }
				\hline \textbf{DS} & \textbf{Metric}  & SSK & CliZZ & SLPA & Walktrap & InfoMap \\
				\hline \parbox[t]{2mm}{\multirow{2}{*}{\rotatebox[origin=c]{90}{LF}}}
				
				 & MSE & 0.026& 0.026 & 0.027& 0.086& 0.035\\

				\cline{2-7}	& RMSE     & 0.161  & 0.0161 & 0.0164 & 0.191& 0.187\\ 
			
				\hline
				\parbox[t]{2mm}{\multirow{2}{*}{\rotatebox[origin=c]{90}{OSS}}} 
			        & MSE & 0.045 & 0.042 & 0.051& 0.047& 0.047\\ 
					\cline{2-7} & RMSE & 0.211   & 0.205 & 0.225 & 0.217 & 0.218\\ 
				
				\hline
				
			\end{tabular}
	
	\label{RegresssionPredictionLinear}
\end{table}


\section{Conclusion and Future Works}
\label{Conclusions}
In this paper, we considered two domains named open source software development and learning networks. We have investigated cooperation \& defection problem on various OSS and LF networks as well as implicit community structures detected by different algorithms. We conclude that OSS networks possess a much higher amount of cooperativity compared to LF networks.We attribute this to the nature of these networks, which in OSS networks, developers cooperate to deliver a product. However, there exist some competency among learners in learning forums, which might lower their cooperation. We as well figured out that cooperation has negative correlation with the network properties. 

Concerning community cooperativity, we conclude that the performance of cooperativity prediction varies based on several aspects. First, we could observe that usage of different community detection algorithms affected on the importance of features. Community detection algorithms which detect high numbers of communities, i.e., small communities, have the density as the most important property. However, community detection which detects lower numbers of communities, i.e., bigger communities, show to have size and average degree as the important features for predicting cooperativity of communities. Second, Spearman correlation values could reveal us traces to see if a property can be important or not. Usually, features with the highest positive or the most negative correlation values showed to be important for predicting the cooperativity of communities. Last but not least, the context of the domains and the used networks are important in such an analysis. In other words, we could figure out that OSS networks have very much lower values for the properties compared to the LF networks. Besides, the amount of cooperativity in OSS communities was higher than LF forums. The changes in the structural formation of the networks of these two domains may cause some variations in our analysis results, which seems reasonable because we constructed different networks with various properties. 

Although in this work we have addressed cooperativity of community structures and their predictions, we plan to investigate the following issues in future studies:
\begin{itemize}
\item We have tested our models on implicit community structures detected by disjoint and overlapping community structures. We could identify algorithms with similar properties concerning community cooperativity, and as well we could specify the best performing algorithms for the prediction models; however, we plan to investigate cooperation \& defection on explicit community structures.
\item We only applied structural properties for the prediction of network and community cooperativity, and we will consider contextual properties of communities like sentiment to investigate their cooperativity. 
\item We intend to apply our models on real environments to figure out how good they perform in real settings. In other words, we can run these models to estimate cooperativity of communities in real settings, e.g., in network of a university.  

\item In addition, we only explored OSS and LF networks for the analysis, and we will apply to rerun our models on large-scale complex networks as well as synthetic networks with community structures.
\end{itemize}

\bibliographystyle{plos2015}
\bibliography{references}
\end{document}